%Paper: hep-th/9509018
%From: sonoda@physics.ucla.edu (Sonoda)
%Date: Tue, 5 Sep 95 12:25:30 -0700

\input harvmac
\input epsf
\noblackbox
\Title{\vbox{\baselineskip 12pt\hbox{UCLA/95/TEP/31}\hbox{hep-th/9509018}}}
{The Energy-Momentum Tensor in Field
Theory II $^\star$
\footnote{}{$^\star$ This work was supported in part
by the U.S. Department of Energy, under Contract
DE-AT03-88ER 40384 Mod A006 Task C.}}

\centerline{
Hidenori SONODA$^\dagger$\footnote{}{$^\dagger$
sonoda@physics.ucla.edu}}
\bigskip\centerline{\it Department of Physics and Astronomy,
UCLA, Los Angeles, CA 90095-1547, USA}

\vskip 1in

In a previous paper, field theory in curved space was
considered, and a formula that expresses the first order
variation of correlation functions with
respect to the external metric was postulated.  The formula
is given as an integral of the energy-momentum tensor over space,
where the short distance singularities of the product of the
energy-momentum tensor and an arbitrary composite field
must be subtracted, and finite counterterms must be added.
These finite counterterms have been interpreted
geometrically as a connection
for the linear space of composite fields over theory space.
In this paper we will study a second
order consistency condition for the variational formula
and determine the torsion of the connection.  A non-vanishing
torsion results from the integrability of the variational formula,
and it is related to the Bose symmetry of
the product of two energy-momentum tensors.
The massive Ising model on a curved two-dimensional surface
is discussed as an example,
and the short-distance singularities of the
product of two energy-momentum tensors are calculated explicitly.

%\draft
\Date{September 1995}

% begin definitions
\def\H{\Theta}
\def\p{\partial}
\def\L{{\cal L}}
\def\mn{{\mu\nu}}
\def\ab{{\alpha\beta}}
\def\dt{{d \over dt}~}
\def\e{{\rm e}}
\def\ep{\epsilon}
\def\vev#1{\left\langle #1 \right\rangle_{h,g}}
\def\vvev#1{\left\langle #1 \right\rangle}
\def\K{{\cal K}}
\def\C{{\cal C}}
\def\Ct{\tilde{\cal C}}
\def\dO{d^{D-1} \Omega}
\def\O{{\cal O}}
\def\gone{g_{\bf 1}}
\def\d{\delta}
\def\bone{\beta_{\bf 1}}
\def\cmmone{(c_m)_m^{~\bf 1}}
\def\det{{\rm det}}
\def\zbar{{\bar z}}

% end definitions

\newsec{Introduction and review}

In a previous paper \ref\rI{H.~Sonoda, ``The Energy-Momentum Tensor
in Field Theory I,'' UCLA preprint (April, 1995),
UCLA/95/TEP/10, hep-th 9504133} we have studied field theory in
curved space and introduced the energy-momentum
tensor as a composite field that generates infinitesimal
deformations of the external metric. It is convenient
to study field theory in an arbitrary curved background
because the energy-momentum tensor is most naturally defined as the field
which is conjugate to the external metric
\ref\rschwinger{J.~Schwinger, Phys.~Rev.~{\bf 127}(1962)324;
{\bf 130}(1962)406, 800}.  The properties of
the energy-momentum tensor in flat space can be obtained by
taking the flat metric limit.  We have introduced
a formula, called the variational formula, which expresses
the first order change of correlation functions of arbitrary
composite fields under an
infinitesimal change of the external metric.
The variational formula treats
the short-distance singularities
in the product of the energy-momentum tensor and an arbitrary
composite field more carefully than earlier studies.\foot{The
necessity of careful treatment of short-distance singularities
in defining integrals over composite fieds was first emphasized
by K.~Wilson \ref\rwilson{K.~G.~Wilson, Phys.~Rev. {\bf D2}
(1970)1478}.}  The singularities
must be subtracted, and finite counterterms,
denoted by $\K$, must be added in the variational formula.
The finite counterterms $\K$ can be interpreted geometrically
as a connection for the linear space of composite fields over
theory space.  (The theory space is parameterized by the external
metric $h_\mn$ and spatially constant parameters $g^i$.)
We have studied the consistency of the variational formula with the
renormalization group (RG), the variational formula with respect
to the spatially constant parameters, and diffeomorphism.
We have obtained two main
results: First, we have obtained an expression
of the short distance singularities of the product of the
energy-momentum tensor and an arbitrary composite field in terms
of the connection $\K$ or finite counterterms.
Second, we have found that the connection $\K$ also gives
the Schwinger terms in the euclidean version of the equal-time
commutator between the energy-momentum tensor
and an arbitrary composite field.

In the present paper we wish to check further consistency
of the variational formula with respect to the
external metric: we will study the second order
variation of the vacuum energy and impose Maxwell's
integrability condition.
This integrability condition is related to the
symmetry of the operator product
expansions (OPE's) under interchange of two Bosonic fields.  We will see
that the integrability condition gives rise to a non-vanishing torsion
of the connection $\K$.

There are two kinds of variational formula for field theory
in D-dimensional curved space with metric $h_\mn$ and
spatially constant parameters $g^i (i=1, ..., N)$.
The first kind, introduced in ref.~\rI, expresses the first
order variation of correlation functions under an arbitrary
change of the external metric $h_\mn$:
\eqn\evar{\eqalign{&\vev{\Phi_{a_1} (P_1) ... \Phi_{a_n} (P_n)}
- \vvev{\Phi_{a_1} (P_1) ... \Phi_{a_n} (P_n)}_{h + \delta h, g}\cr
& = \lim_{\ep \to 0} \Bigg[
\int_{\rho (r,P_k) \ge \epsilon} d^D r \sqrt{h} ~{1 \over 2} \delta
h_\mn (r) \cr
& \qquad \qquad \qquad
\times \vev{ \left( \H^\mn (r) - \vev{\H^\mn (r)} \right)
\Phi_{a_1} (P_1) ... \Phi_{a_n} (P_n)} \cr
& + \sum_{k=1}^n
\left\{ \d h \cdot \K (h(P_k),g) -
\int_\ep^1 d\rho~ \d h \cdot \C (\rho; h(P_k),g) \right\}_{a_k}^{~b}\cr
&\qquad\qquad \times
\vev{\Phi_{a_1} (P_1) ...  \Phi_b (P_k) ... \Phi_{a_n} (P_n)}
\quad\Bigg]~,\cr}}
where $\rho (r,P)$ is the geodesic distance between the
two points $r$ and $P$.
The OPE coefficient $\C$ is defined by
\eqn\eC{\int_{\rho (r,P) = \rho}
\dO~ {1 \over 2} \d h_\mn \H^\mn (r) \Phi_a (P)
= \left[ \d h \cdot \C (\rho;g,h(P) \right]_a^{~b}
\Phi_b (P) + {\rm o} \left( {1 \over \rho} \right)~,}
where $\dO$ is the angular volume element:
\eqn\edO{d^D r \sqrt{h} = d\rho~ \dO~,}
and $\d h \cdot \C$ is a short-hand notation for
\eqn\eCnotation{\left[ \d h \cdot \C (\rho;g,h)\right]_a^{~b}
\equiv \sum_{m=0}^\infty
{1 \over m!}~\nabla_{\mu_1} ... \nabla_{\mu_m}
{1 \over 2}~\d h_\mn \cdot \left( \C^{\mn, \mu_1 ... \mu_m} (\rho;g,h)
\right)_a^{~b}~.}
In eq.~\eC\ we only keep the terms which cannot be integrated
over $\rho$ all the way to zero.  The definition \eC\ implies
that the OPE coefficients satisfy the algebraic constraints
\eqn\eCconstraint{\C^{\mn, \mu_1 ... \mu_m} (\rho;h,g)
= {1 \over \rho^2} h_{\mu_{m+1} \mu_{m+2}}
\C^{\mn, \mu_1 ... \mu_{m+1} \mu_{m+2}} (\rho;h,g) + {\rm o} \left( {1 \over
\rho^3} \right)~.}

In ref.~\rI\ we have imposed the consistency of
the variational formula \evar\ with the RG, and found the
following expression of the OPE coefficient
$\d h \cdot \C$ in terms of the connection $\K$:
\eqn\eCdS{\d h \cdot
\C (\rho ;h,g) = {\p \over \p \rho}~\d h \cdot S(\rho; h,g)~,}
where the matrix $S(\rho;h,g)$ is defined by
\eqn\eS{\eqalign{\delta h \cdot S(\rho; h,g) &\equiv
\Bigg[~ G(\rho; h,g) \cdot \left\{
{\delta h \over \rho^2} \cdot
\K \left( h/\rho^2, g(\ln \rho) \right) \right\}\cr
&\quad +
G(\rho; h,g) - G(\rho; h+\delta h,g) ~\Bigg]
\cdot G^{-1} (\rho;h,g)~.\cr}}
The matrix $G$ is defined by the following RG equation and
the initial condition:
\eqn\eG{\eqalign{\dt G (\rho; h,g) &\equiv
\lim_{\Delta t \to 0} {1 \over \Delta t}
\left[ G(\e^{- \Delta t} \rho; \e^{- 2 \Delta t} h, g
+ \Delta t \beta) - G(\rho; h,g) \right] \cr
& = \gamma (h,g) G(\rho; h,g)~,
\quad G(1;h,g) = {\bf 1}~,\cr}}
where $\gamma (h,g)$ is the matrix of scale dimension
in the basis $\{\Phi_a\}$,
and the running parameter $g^i (t)$ is defined by
\eqn\erunning{{\p \over \p t}~ g^i (t) = \beta^i (g(t))~,\quad
g^i (0) = g^i~,}
where $\beta^i (g)$ is the beta function of the $i$-th
parameter $g^i$.  Note that the initial condition for $G$
implies
\eqn\eSK{S(\rho = 1; h,g) = \K (h,g)~.}

Similarly, the consistency with diffeomorphism gives rise to
the following expression of the euclidean version of
the equal-time commutator between the energy-momentum tensor and
an arbitrary composite field $\Phi_a$:
\eqn\ecomm{(u \cdot \Ct (\rho; h,g)) \Phi_a (P)
= \L_u \Phi_a + \left( \L_u h \cdot
S (\rho; h, g) \right) \Phi_a ,}
where $u$ is an arbitrary vector field, $\L_u$ is
the Lie derivative in the direction of $u$, and
$\Ct$ is defined by
\eqn\eCt{\int_{\rho (r,P) = \rho} \dO (r)~N_\mu (r) u_\nu (r)
\H^\mn (r) \Phi_a (P) = (u \cdot \Ct (\rho; h,g)) \Phi_a (P)
+ {\rm o} \left( \rho^0 \right)~,}
where $N^\mu (r)$ is the unit outward normal vector at $r$,
and we only keep the terms non-vanishing as $\rho \to 0$.
Eq.~\ecomm\ shows that the anomalous part of the commutator is
determined by the connection $\K$ through $S$ defined by
eq.~\eS.  In components, $\Ct$ is written as
\eqn\eCtnotation{(u \cdot \Ct)_a^{~b} (\rho;h,g)
= \sum_{m=0}^\infty {1 \over m!}~\nabla_{\mu_1}
... \nabla_{\mu_m} u_\nu \cdot (\Ct^{\mn,~\mu_1
... \mu_m}_{~~~~\mu})_a^{~b} (\rho; h,g)~,}
where
\eqn\eCttrace{\Ct^{\mn,~\mu_1
... \mu_m}_{~~~~\mu} (\rho;h,g) = h_{\mu\mu_{m+1}}
\Ct^{\mn,\mu_{m+1} \mu_1 ... \mu_m} (\rho;h,g)~,}
and the unintegrable part of $\rho \Ct$'s coincide
with $\C$'s:
\eqn\eCCt{\rho ~\Ct^{\mn,\mu_1 ... \mu_m} (\rho; h,g)
= \C^{\mn,\mu_1 ... \mu_m} (\rho; h,g) +
{\rm o} \left( {1 \over \rho} \right)~.}

There are four sources of algebraic constraints
on the connection $\K$.  One is the constraint
\eCconstraint.  Since Eq.~\ecomm\ determines $\Ct^{\mn,~\mu_1
... \mu_m}_{~~~~\mu}$, which is a trace over $\mu$,
the existence of $\Ct^{\mn,\mu_1 ... \mu_m}$ which
gives the correct trace \eCttrace\
can constrain the connection $\K$.  This is the second
constraint.  The relation \eCCt\ can give the
third constraint, since both $\C$'s and $\Ct$'s are given in terms of
the same connection $\K$.  Finally, algebraic constraints
analogous to \eCconstraint\ exist also for $\Ct$'s:
\eqn\eCtconstraint{\Ct^{\mn, ~~\mu_1 ... \mu_m}_{~~~~\mu} (\rho)
= {1 \over \rho^2} h_{\mu_{m+1} \mu_{m+2}}
\Ct^{\mn, ~~\mu_1 ... \mu_{m+2}}_{~~~~\mu} (\rho) + {\rm o}
\left({1 \over \rho^2}\right)~.}
These constrains are very useful when we determine
the connection $\K$ in practice.

The second kind of variational formula gives the first order
change of correlation functions under an arbitrary change of
the parameter $g^i$ of the theory
\ref\rconn{H.~Sonoda, {\it in} Proceedings of the Conference
Strings '93, eds. M.~B.~Halpern, G.~Rivlis, and A.~Sevrin (World
Scientific, 1995), and references therein}:
\eqn\evarpar{\eqalign{& - {\partial \over \partial g^i}
\vev{\Phi_{a_1} (P_1) ... \Phi_{a_n} (P_n)} \cr
& = \lim_{\ep \to 0} \Bigg[ \int_{\rho (r,P_k) \ge \ep} d^D r \sqrt{h}~\vev{
\left( \O_i (r) - \vev{\O_i (r)} \right)
\Phi_{a_1} (P_1) ... \Phi_{a_n} (P_n)} \cr
&\qquad + \sum_{k=1}^n
\left\{ (c_i) (h(P_k),g) - \int_\ep^1
d\rho~(\C_i) (\rho; h(P_k),g) \right\}_{a_k}^{~b}\cr
&\qquad\qquad\qquad \times
\vev{\Phi_{a_1} (P_1) ... \Phi_b (P_k) ... \Phi_{a_n} (P_n)} \Bigg]~,\cr}}
where
\eqn\eCi{\int_{\rho (r,P) = \rho}
\dO~ \O_i (r) \Phi_a (P) = \left[ \C_i (\rho;g,h(P)) \right]_a^{~b}
\Phi_b (P) + {\rm o} \left( {1 \over \rho} \right)~,}
and the finite counterterm $c_i$ can be interpreted
as the $g^i$ component of a connection over theory space.

The consistency between the variational formula \evarpar\
and the RG gives the following relation between the
OPE coefficient $\C_i$ and the connection $c_i$ \rconn:
\eqn\eCic{\C_i (\rho ;h,g) =
{\p \over \p \rho} S_i (\rho; h,g)~,}
where
\eqn\eSi{S_i (\rho; h,g) \equiv \left[
G (\rho; h,g) {\p g^j (\ln \rho) \over \p g^i} ~c_j (h,g)
- {\p \over \p g^i} G (\rho;h,g) \right] \cdot ~G^{-1} (\rho; h,g) ~.}

It has also been found that the consistency among the two kinds of variational
formula and the RG gives the trace condition \rI:
\eqn\etrace{2 h \cdot \K (h,g) = \gamma (h,g) + \beta^i (g)
c_i (h,g)~.}

The purpose of the present paper is to check further the
consistency of the variational formula
with respect to the metric \evar.  In particular
we will study the second order variation of the
vacuum energy using the first order variational formula \evar\
recursively.  We will find that the symmetry of
the second order variation results in a non-vanishing torsion
of the connection $\K$.  It will be shown that part of this result
is equivalent to the Bose symmetry between two energy-momentum tensors.

The paper is organized as follows.  In sect.~2 we will
calculate the second order variation of the vacuum energy
using the variational formula \evar.  By imposing
an integrability condition (or Maxwell's relation) to the
second order variation, we will derive the torsion of the
connection $\K$.  In sect.~3 we will study an implication
of the Bose symmetry among the composite fields with integer
spins, and show that part of the result of sect.~2 can be also obtained
from the Bose symmetry between two energy-momentum tensors.
In sect.~4 we will find a relation between the particular
matrix elements of the connections $c_i$ and $\K$, i.e.,
$(c_i) \H^\mn$ and $\K \O_i$.  In sect.~5 we will study the example
of the massive Ising model on a curved two-dimensional
surface, and compute explicitly $\K \H^\mn$, i.e.,
the elements of the connection $\K$ for the energy-momentum tensor.
In the zero mass limit, the result is shown to agree with
conformal field theory.  We conclude the paper in sect.~6.

\newsec{Second order variation of the vacuum energy}

We consider a field theory in $D$-dimensional curved space
with an external metric $h_\mn$ and spatially constant
parameters $g^i (i=1, ..., N)$.
Let $F[h,g]$ be the total vacuum energy (or free energy).
Under an infinitesimal change of the metric and the parameters,
the free energy changes by
\eqn\efirstorder{\eqalign{
&F[h + \d h, g + \d g] - F[h,g]\cr
& = \int d^D P \sqrt{h (P)}\;\left[
{1 \over 2} \d h_\mn (P) \vev{\H^\mn (P)}
+ \d g^i \vev{\O_i (P)} \right]\;.\cr}}
We wish to calculate the terms of order $\d h_1 \d h_2$ in
\eqn\esecond{\Delta [\d h_1, \d h_2] \equiv F[h + \d h_1 + \d h_2,g]
- F[h+ \d h_1,g] - F[h + \d h_2,g] + F[h,g]~.}
(We ignore the terms of order $(\d h_1)^2$ and $(\d h_2)^2$.)
The integrability of the first order variational formula
\efirstorder\ demands that
the second order variation \esecond\ must be symmetric with
respect to $\d h_1$ and $\d h_2$.

We can calculate the second order change \esecond\
by applying the variational formula \evar\ to eq.~\efirstorder.
There are two ways of doing this depending on the order
of changing the metric, as indicated in Fig. 1.  Integrability
of the vacuum energy demands that the two paths in the figure
give the same result.
\vskip .2in
\centerline{\epsfxsize = 0.6\hsize \epsfbox{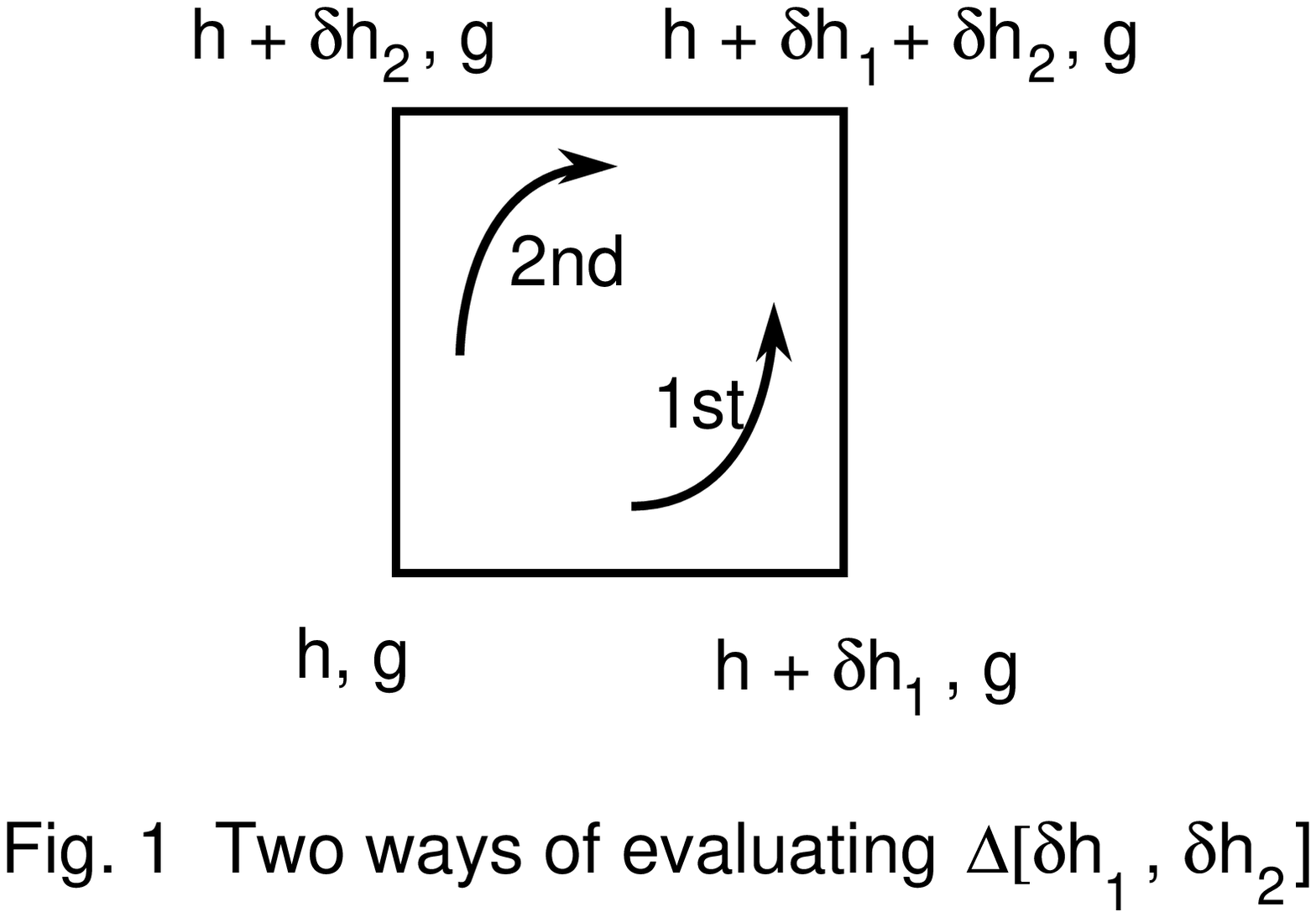}}

\noindent
Along the first path, we obtain
\eqn\estepone{\eqalign{
\Delta  [\d h_1, \d h_2]
&= ( F[(h + \d h_1) + \d h_2, g] - F[h+\d h_1,g] )\cr
&\qquad\qquad - ( F[h + \d h_2, g] - F[h,g] )\cr
&= \int d^D P \sqrt{h + \d h_1}\;{1 \over 2} (\d h_2)_\mn (P)
\vvev{\H^\mn (P)}_{h + \d h_1, g}\cr
&\qquad\qquad - ( F[h + \d h_2, g] - F[h,g] )\;.\cr}}
Applying the variational formula \evar\ to the integrand, we obtain
\eqnn\esecondorder
$$
\eqalignno{&\Delta  [\d h_1, \d h_2]
=  - \int_{\rho (r,P) \ge \ep} d^D P \sqrt{h(P)}~d^D r
\sqrt{h(r)}\cr
&\qquad \times {1 \over 2} (\d h_2)_\mn (P) {1 \over 2}
(\d h_1)_\ab (r) \vev{\H^{\alpha \beta} (r) \H^\mn (P)}^c\cr
& ~+ \int d^D P \sqrt{h (P)} ~{1 \over 2}
(\d h_2)_\mn (P) \Bigg\{
h^\ab (P) {1 \over 2} (\d h_1 (P))_{\alpha\beta}
\vev{\H^\mn (P)} &\esecondorder\cr
&\qquad + \d h_1 (P) \cdot
\left( \int_\ep^1 d\rho~\C (\rho; h(P),g)
- \K (h(P),g) \right)^{\mn\; a} \vev{\Phi_a (P)} \Bigg\}~.\cr}
$$
The second path gives the same result
as above except that $\d h_1$ and $\d h_2$
are interchanged.

Since the energy-momentum tensor satisfies the canonical
RG equation
\eqn\eRGH{\dt \H^\mn = (D+2) \H^\mn~,}
the general formulas \eCdS, \eS\ imply
\eqn\eSH{\d h \cdot \C (\rho;h,g) \H^\mn
= {\p \over \p \rho} \left( \d h \cdot S (\rho;h,g) \right)
\H^\mn~,}
where
\eqn\eSKH{\d h \cdot S (\rho;h,g) \H^\mn
= {1 \over \rho^{D+2}} {\d h \over \rho^2} \cdot
\K \left( h/\rho^2, g(\ln \rho) \right)
\cdot G^{-1} (\rho; h,g) \H^\mn~.}
Using the component notation of eq.~\eCnotation,
this can be written as
\eqn\eSHcomp{\left[ \C^{\ab, \mu_1 ... \mu_m}
(\rho;h,g) \right]^{\mn~a}
= {\p \over \p \rho} \left[ S^{\ab, \mu_1 ... \mu_m} (\rho;h,g)
\right]^{\mn~a}~,}
where
\eqn\eSKcomp{\eqalign{&\left[ S^{\ab, \mu_1 ... \mu_m}
(\rho; h,g) \right]^{\mn~a}\cr
&\quad = {1 \over \rho^{D+4}} \left(
\K^{\ab, \mu_1 ... \mu_m} \left(
h/ \rho^2, g(\ln \rho) \right) \right)^{\mn~b} \cdot
(G^{-1})_b^{~a} (\rho; h,g)~.\cr}}

Therefore, the invariance of eq.~\esecondorder\
under the interchange of $\d h_1$ and $\d h_2$ implies that
\eqn\emaxwell{I_{12} (\ep) = I_{21} (\ep)~,}
where
\eqn\eIdef{\eqalign{I_{12} (\rho) &\equiv
\int d^D P \sqrt{h(P)} {1 \over 2} (\d h_2)_\mn (P)
\Bigg( - {1 \over 2} (\d h_1)_\ab h^\ab
\H^\mn(P) \cr
+ & \sum_{m=0}^\infty {1 \over m!} \nabla_{\mu_1}
... \nabla_{\mu_m} {1 \over 2} (\d h_1)_\ab (P) \cdot
S^{\ab, \mu_1 ... \mu_m} ( \rho; h,g) \H^\mn \Bigg) ~.\cr}}
We recall that the connection
$\K$ was originally introduced in ref.~\rI\ as finite
counterterms in the variational formula \evar.
This means that in the definition of $I_{12} (\rho)$
above, $S (\rho)$ only has those terms which give
non-vanishing contributions for an infinitesimal
$\rho$.  This implies that eq.~\emaxwell\ is actually valid for
an arbitrary $\ep$ which is not necessarily infinitesimal
as long as $I_{12} (\ep)$ is well-defined.

Therefore, from eq.~\emaxwell, we find that
\eqn\etotaldiv{\eqalign{f(\rho) &\equiv (\d h_1)_\mn \Big[ h^\mn
(\d h_2)_\ab \H^\ab \cr
&~+ \sum_{m=0}^\infty {1 \over m!}~\nabla_{\mu_1}
... \nabla_{\mu_m} (\d h_2)_\ab \cdot
S^{\ab, \mu_1 ... \mu_m} ( \rho; h,g) \H^\mn \Big]
- ( \d h_1 \leftrightarrow \d h_2 ) \cr}}
is a total divergence with respect to space.  Since
$\d h_1$, $\d h_2$ are arbitrary, this means that for each
integer $m$ we must find
\eqn\eintegrability{\eqalign{
& S^{\ab, \mu_1 ... \mu_m}
(h/\rho^2, g(\ln \rho)) \H^\mn \cr
& = \d_{m,0} ~(h^\ab \H^\mn - h^\mn \H^\ab) \cr
& \quad + (-)^m
\sum_{n=0}^\infty {(-)^n \over n!}~\nabla_{\nu_1}
... \nabla_{\nu_n} \left\{ S^{\mn, \mu_1 ... \mu_m \nu_1
... \nu_n} (h/\rho^2, g(\ln \rho)) \H^\ab \right\} ~.\cr}}
This is the main result of this section.
The first term on the right-hand side is independent of
the geodesic distance $\rho$.  In the next section we will
see that the $\rho$ dependence of eq.~\eintegrability\ is
a consequence of the Bose statistics of the energy-momentum
tensor.

Finally we observe that, if we take $\rho =1$,
eq.~\eintegrability\ gives the torsion of the
connection $\K$:
\eqn\etorsion{\eqalign{&\tau (\d h_1, \d h_2;h,g) \equiv
(\d h_1 \cdot \K) (\d h_2)_\ab \H^\ab - (\d h_2 \cdot \K)
(\d h_1)_\ab \H^\ab \cr
& = {1 \over 2} \left\{ (\d h_1)_\mu^\mu  (\d h_2)_\ab \H^\ab
- (\d h_2)_\mu^\mu (\d h_1)_\ab \H^\ab \right\} \cr
+& \sum_{m=1}^\infty {1 \over m!}~ \Bigg[
(\d h_2)_\ab (-)^m \nabla_{\mu_1} ... \nabla_{\mu_m}
\left( {1 \over 2} (\d h_1)_\mn \cdot
(\K^{\ab, \mu_1 ... \mu_m}) \H^\mn \right) \cr
&\quad - \left( \nabla_{\mu_1} ... \nabla_{\mu_m} (\d h_2)_\ab \right) \cdot
{1 \over 2} (\d h_1)_\mn (\K^{\ab, \mu_1 ... \mu_m}) \H^\mn \Bigg]~.\cr}}
The sum over the positive integer $m$ is a total derivative in space.

To summarize, the condition of integrability of the vacuum energy gives
eq.~\eintegrability. This determines the torsion of the
connection $\K$ up to total derivatives in space.

\newsec{Bose symmetry of the OPE coefficients in curved space}

To improve our understanding of the integrability condition
\eintegrability\ in the previous section, let us study the
property of OPE's in curved space under interchange of two
fields.

In curved space we consider an OPE of two composite fields
\eqn\eopeab{A (r) B (P) = C_{AB}^{~~a} (P,v) \Phi_a (P) \quad,}
where $v$ is the geodesic coordinate of $r$ with respect to $P$
(i.e., $r = {\rm Exp}_v (P)$).
For simplicity, we take $A$ and $B$ to be scalar fields.
We define the angular integral
\eqn\ecab{\C_{AB}^{\mu_1 ... \mu_m, a} (\rho; h(P)) \equiv
\int_{\| v\| = \rho} \dO_\rho (r) v^{\mu_1} ... v^{\mu_m}
C_{AB}^{~~a} (P,v)~,}
where $d^D r \sqrt{h} = d\rho \dO (r)$, and $\|v \|$
is the norm.
Our goal is to find the symmetry of the integrated OPE
coefficient under interchange of $A$ and $B$: we wish to know
the relation between $\C_{AB}^{\mu_1 ... \mu_m, a}$ and
$\C_{BA}^{\mu_1 ... \mu_m, a}$.

Let us introduce $V^a_{~b} (P,r)$ which parallel transports
a tensor $\Phi_a$ at $P$ to a tensor at
$r = {\rm Exp}_v (P)$ along the geodesic.  The transported
tensor is given by $\sum_b \Phi_b (P) V^b_{~c} (P,r)$.  The
inverse of $V$ parallel transports a tensor from $r$ to $P$.
So, we can write
\eqn\eVinverse{\left( V (P,r) \right)^{-1} = V (r,P)~.}

Now, the Bose symmetry
\eqn\ebose{B(r) A(P) = A(P) B(r)}
implies that
\eqn\eopebose{B(r) A(P) = \sum_a
C_{AB}^{~~a} (r, w) \Phi_a (r) ,}
where $w$ is the tangent vector at $r$ such that ${\rm Exp}_w (r) = P$.
(See Fig. 2.)
\vskip .2in
\centerline{\epsfxsize=0.4\hsize \epsfbox{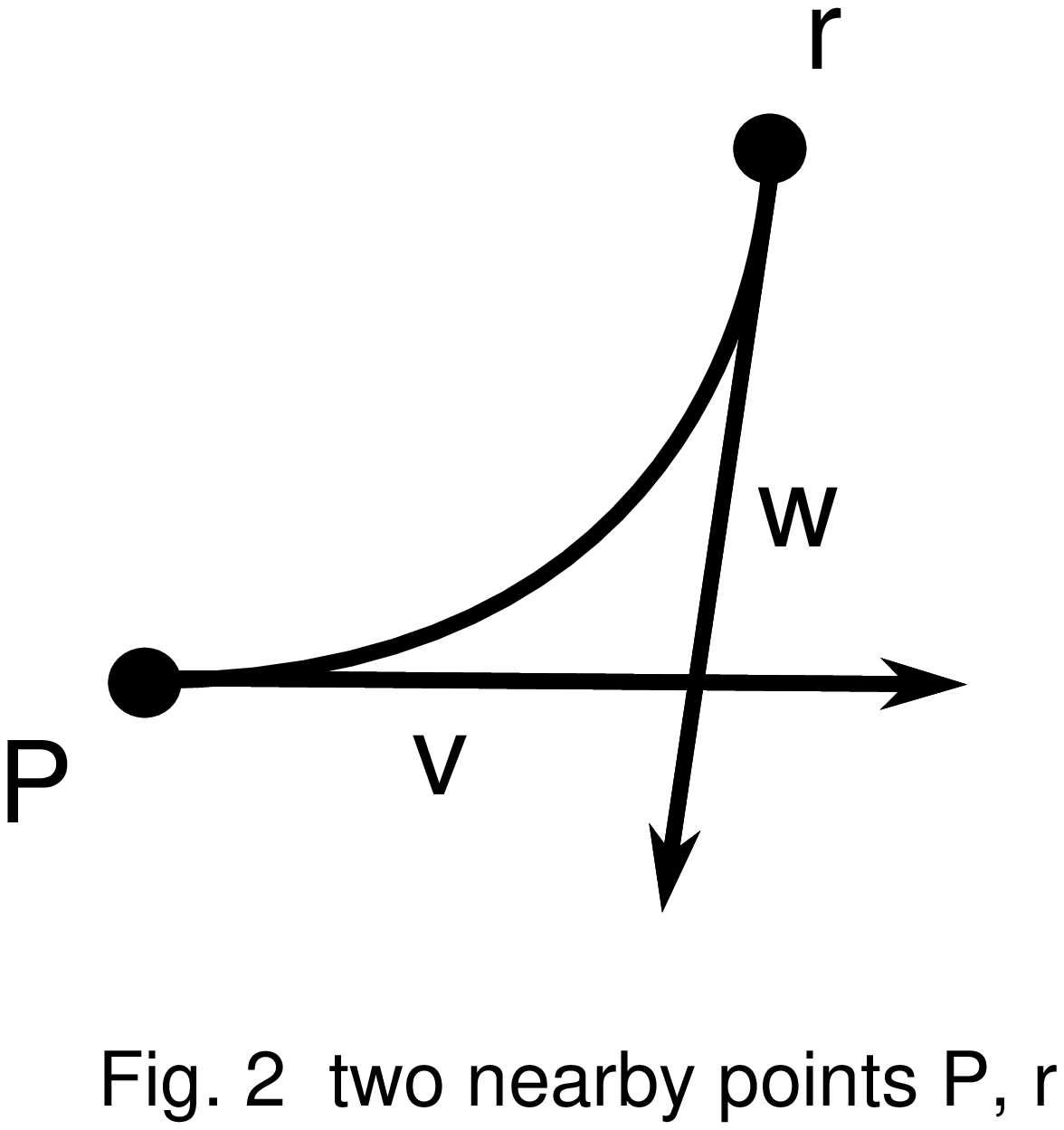}}

\noindent
Using the Taylor expansion
\eqn\etaylor{\Phi_a (r)
= \sum_{n=0}^\infty {1 \over n!}~v^{\nu_1} ... v^{\nu_n}
\nabla_{\nu_1} ... \nabla_{\nu_n} \Phi_b (P) \cdot V^b_{~a} (P,r)~,}
we obtain
\eqn\ebegin{B (r) A (P) = (C_{AB})^a (r,w)
\sum_{n=0}^\infty {1 \over n!}~v^{\nu_1} ... v^{\nu_n}
\nabla_{\nu_1} ... \nabla_{\nu_n} \Phi_b (P) \cdot V^b_{~a} (P,r)~.}
Therefore, using the definition \ecab, we find
\eqn\eintermedi{\eqalign{&\C_{BA}^{\mu_1 ... \mu_m, a}
(\rho; h,g) \Phi_a (P)\cr
=& \sum_{n=0}^\infty {(-)^{m+n} \over n!}
\int \dO_\rho (v) (-v)^{\mu_1} ... (-v)^{\mu_m}
(-v)^{\nu_1} ... (-v)^{\nu_n} \cr
&\qquad\qquad \times V^b_{~a} (P,r) C_{AB}^{~~a}
(r,w) \nabla_{\nu_1} ... \nabla_{\nu_n} \Phi_b (P)~.\cr}}

In order to proceed further we need three equations.
The first equation is the Taylor expansion
\eqn\efirst{V^b_{~a} (P,r) C_{AB}^{~~a} (r,w)
= \sum_{k=0}^\infty {1 \over k!}~v^{\kappa_1} ... v^{\kappa_k}
\nabla_{\kappa_1} ... \nabla_{\kappa_k} C_{AB}^{~~b} (P, -v)~.}
This is valid since we can formally expand the OPE coefficients
as
\eqn\eexpand{C_{AB}^{~~a} (P,v) = \sum_{m=0}^\infty
{1 \over m!}~v^{\mu_1} ... v^{\mu_m} C_{AB, \mu_1
... \mu_m}^{~~a} (P;\Vert v \Vert)~,}
where $C_{AB, \mu_1 ... \mu_m}^{~~a} (P;\Vert v \Vert)$,
except for its dependence on the norm $\Vert v \Vert$, is an
ordinary tensor field and admits the Taylor expansion
\eqn\etaylorCAB{\eqalign{& V^a_{~b} (P,r)
C_{AB, \mu'_1 ... \mu'_m}^{~~b} (r;\Vert v
\Vert) V^{\mu'_1}_{~\mu_1} (r,P) ... V^{\mu'_m}_{~\mu_m} (r,P) \cr
& = \sum_{n=0}^\infty {1 \over n!}~v^{\nu_1} ... v^{\nu_n}~\nabla_{\nu_1}
... \nabla_{\nu_n}  C_{AB, \mu_1 ... \mu_m}^{~~a} (P;\Vert v \Vert) ~,\cr}}
just as eq.~\etaylor.

To describe the other two equations, we
first introduce $W^\mu_{~\nu} (P,v)$ which parallel transports
tangent vectors from $r = {\rm Exp}_v (P)$ to $P$ in the
geodesic coordinate system around $P$.  The two
parallel transports $V$ and $W$ are
related to each other by the following coordinate transformation:
\eqn\eWV{W^\mu_{~\nu} (P,v) = V^\mu_{~\alpha} (P,r) {\p v_r^\alpha
\over \p v_P^\nu}\bigg|_{v_r = 0}~,}
where the vector $v_P$ at $P$ and the vector $v_r$ at $r$
correspond to the same point: ${\rm Exp}_{v_P} (P) = {\rm Exp}_{v_r} (r)$.
In the geodesic coordinate $v_P$ around $P$, the volume element
at the point ${\rm Exp}_{v_P} (P)$ is given by $\det W(P,v_P)$.
The second equation we need is the following expansion of
the volume element:
\eqn\esecond{\det W (P,v) = \sum_{k=0}^\infty
{1 \over k!}~v^{\kappa_1} ... v^{\kappa_k}
\nabla_{\kappa_1} ... \nabla_{\kappa_k} \det W(P,-v)~.}
The third equation we need is the one about integrals of
total derivatives:
\eqn\ethird{\eqalign{&\nabla_{\mu_1} ... \nabla_{\mu_m}
\int \dO_\rho (v)~v^{\mu_1} ... v^{\mu_m} t (P,v)\cr
=& \int {\dO_\rho (v) \over \det W(P,v)}~v^{\mu_1} ...
v^{\mu_m} \nabla_{\mu_1} ... \nabla_{\mu_m}
\left( \det W(P,v) t (P,v) \right)~,\cr}}
where $t(P,v)$ is an arbitrary tensor at $P$ which depends also
on a vector $v$ at $P$.  The tensor can be formally
expanded as
\eqn\etpvtaylor{t(P,v) = \sum_{n=0}^{\infty}
{1 \over n!}~v^{\nu_1} ... v^{\nu_n}
t_{\nu_1 ... \nu_n} (P, \Vert v \Vert) ~,}
where $t_{\nu_1 ... \nu_n} (P, \Vert v \Vert)$ is an ordinary
tensor field with an additional dependence on the norm
of the vector $v$.  The covariant derivative of $t(P,v)$
is defined by
\eqn\edefderiv{\nabla_\mu t (P,v) \equiv
\sum_{m=0}^\infty {1 \over n!}~v^{\nu_1} ... v^{\nu_n}~\nabla_\mu
t_{\nu_1 ... \nu_n} (P,\Vert v \Vert)~,}
where we have an ordinary covariant derivative on the
right-hand side.  ($\Vert v \Vert$ is fixed under the
derivative.)
We will not prove the second and third equations
\esecond, \ethird\ in this paper.  For a mathematical
background we refer the reader to
refs.~\ref\rfriedan{D.~Friedan, Ann.~Phys. {\bf 163}(1985)318}
and \ref\rkn{S.~Kobayashi and K.~Nomizu, Foundations of
Differential Geometry I (Interscience, 1963)}.
($W$ here is denoted as $V$ in these references.)

Using eqs.~\efirst\ and \esecond, eq.~\eintermedi\ gives
\eqn\eintermedii{\eqalign{&\C_{BA}^{\mu_1 ... \mu_m, a}
(\rho; h,g) \Phi_a (P)\cr
&= (-)^m \sum_{n=0}^\infty
{(-)^n \over n!} \int {\dO_\rho (v) \over
\det W (P,v)}~v^{\nu_1} ... v^{\nu_n} \cr
&\qquad \times \nabla_{\nu_1} ... \nabla_{\nu_n} \left( \det W(P,v)
v^{\mu_1} ... v^{\mu_m} C_{AB}^{~~a} (P,v) \Phi_a (P) \right)\cr}}
after a change of coordinates $v \to - v$.
Then, by using eq.~\ethird, we obtain
the desired relation:
\eqn\einterchange{\eqalign{&\C_{BA}^{\mu_1 ... \mu_m, a}
(\rho; h,g) \Phi_a (P) \cr
=& (-)^m \sum_{n=0}^\infty
{(-)^n \over n!}~\nabla_{\nu_1} ... \nabla_{\nu_n}
\left[ \C_{AB}^{\mu_1 ... \mu_m \nu_1 ... \nu_n, a}
(\rho; h,g) \Phi_a (P) \right]~.\cr}}
This implies that the difference
\eqn\ediff{\C_{BA}^{\mu_1 ... \mu_m, a}
(\rho; h,g) \Phi_a (P) - (-)^m \C_{AB}^{\mu_1 ... \mu_m, a}
(\rho; h,g) \Phi_a (P)}
is a total derivative in space.  Eq.~\einterchange\ is
the main result of this section; it is a direct
consequence of the Bose symmetry \ebose.

Now for the energy-momentum tensor, the only difference
is that it is a tensor, and the integrated OPE must be
defined with the parallel transport operator as
\eqn\eintope{\eqalign{&\int_{\rho (r,P) = \rho} \dO (r)~v^{\mu_1} ... v^{\mu_m}
V^\alpha_{~\gamma} (P,r) V^\beta_{~\d} (P,r)
\H^{\gamma\d} (r) \H^\mn (P)\cr
&\qquad = \sum_a \left[\C^{\ab, \mu_1 ... \mu_m}
(\rho; h (P),g) \right]^{\mn~a}
\Phi_a (P)~.\cr}}
If we only keep the terms which cannot be integrated over
$\rho$ to zero, this $\C$ coincides with the $\C$ in the
previous sections.  The relation analogous to eq.~\einterchange\
is given by
\eqn\einterchangeH{\eqalign{
&\left( \C^{\ab, \mu_1 ... \mu_m} (\rho; h,g) \right)^{\mn~a}
\Phi_a (P) \cr
= &
(-)^m \sum_{n=0}^\infty {(-)^n \over n!}~\nabla_{\nu_1}
... \nabla_{\nu_n} \left[
\left( \C^{\mn, \mu_1 ... \mu_m \nu_1 ... \nu_n} (\rho; h,g)
\right)^{\ab~a} \Phi_a (P) \right]~.\cr}}
This is precisely what we obtain by
differentiating eq.~\eintegrability\ with respect to $\rho$,
thanks to the relation \eCdS\ between $\C$ and $S$.  Hence,
the $\rho$ dependence of the integrability condition
\eintegrability\ is consistent with the invariance
of the OPE under interchange of two energy-momentum tensors.

Eq.~\einterchange\ generalizes easily to a product of
any two bosonic composite fields.

\newsec{A relation between $c_i$ and $\K$}

In the same way as in sect.~2, the symmetry of
the second order variation
\eqn\edhdg{\Delta[\d h, \d g] \equiv
F[h + \d h, g + \d g] - F[h + \d h, g] - F[h, g + \d g]
+ F[h,g]}
gives a relation among the mixed matrix elements of the
connection $(c_i, \K)$.
\vskip 0.2in
\centerline{\epsfxsize=0.6\hsize \epsfbox{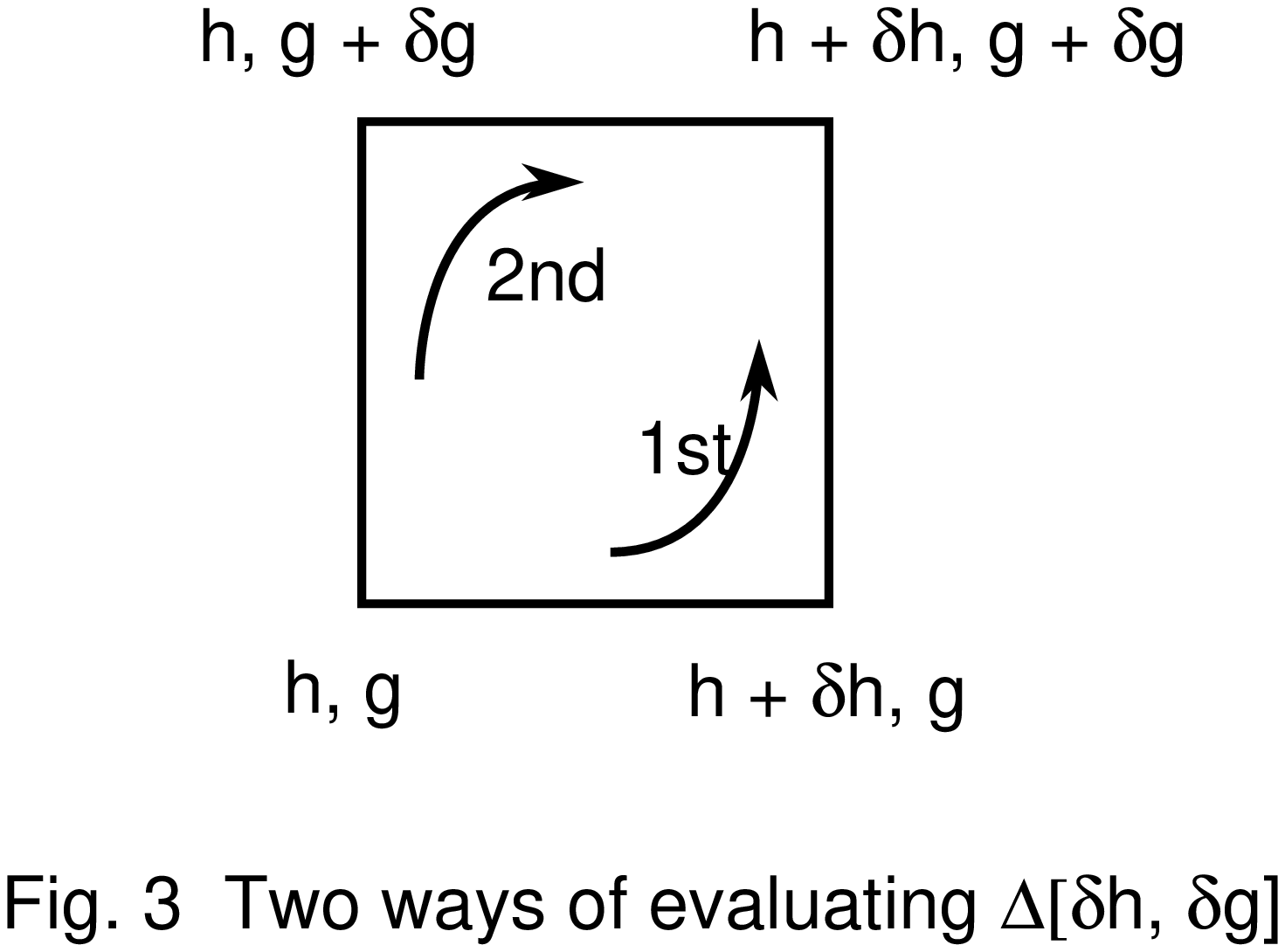}}

\noindent
By demanding that the two
paths shown in Fig.~3 give
the same second order variation \edhdg, we find the following
relation (we skip the derivation, since it is analogous
to the derivation in sect.~2):
\eqn\ecKrelation{(c_i) \H^\mn - (\K^\mn) \O_i
= - h^\mn \O_i +
\sum_{m=1}^\infty {(-)^m \over m!}~\nabla_{\mu_1} ... \nabla_{\mu_m}
\left[ \left( \K^{\mn,~\mu_1 ... \mu_m}\right) \O_i \right]~.}
This also gives torsion of the connection
$(c_i, \K)$, which is related to the Bose symmetry of the
operator product $\O_i (r) \H^\mn (P)$ under interchange of
the two fields.

\newsec{Example: Massive Ising model on a curved surface}

We have obtained two constraints \eintegrability (or
equivalently \etorsion) and \ecKrelation.  As an example
of a practical use of these constraints,
we take the massive Ising model on a curved two-dimensional
surface with metric $h_\mn$, and apply the integrability
constraints and the other algebraic constraints discussed in sect.~1
to determine the connection $\K$ explicitly.

The model has three parameters: $\gone$,
$m$, and $\kappa$, where $\gone$ is the cosmological constant,
$m$ is the mass, and $\kappa$ is the coefficient of the
Ricci curvature $R$ in the vacuum energy density.  The metric and the
parameters satisfy the following RG equations:
\eqn\erg{\eqalign{&\dt h_\mn = 2 h_\mn~, \cr
&\dt \gone = 2 \gone + {m^2 \over 2} \bone~, \quad
\dt m = m~, \quad \dt \kappa = c' ~,\cr}}
where $c'$ is a constant to be determined later.
The constant $\bone$ depends on how we normalize $m$.
If we choose $m$ such that it gives the physical mass
of the free fermion in flat space, then
\eqn\ebone{\bone = - {1 \over 2 \pi}~.}

The above RG equations for the parameters imply
the following RG equation for the field $\O_m$ conjugate
to $m$:
\eqn\ergOm{\dt \O_m = \O_m - m \bone~.}
Eqs.~\erg\ also imply the following trace of the
energy-momentum tensor:
\eqn\eIsingtrace{\H = 2 \gone + {m^2 \over 2} \bone + m \O_m +
c' R~.}
By taking the limit $m=0$, we find that the constant $c'$ is related to
the central charge $c = {1 \over 2}$ of the conformal Ising
model as
\eqn\eccprime{c' = - {c \over 24 \pi}~.}

Our goal is to calculate the singularities in the product
of two energy-momentum tensors $\H^\mn (r) \H^\ab (P)$
explicitly.  The answer is well-known in the massless
limit $m=0$, but in the following we will be able to
calculate the correction due to the non-vanishing mass $m$.

As a preparation we first calculate the
singularities in the product $\H^\mn (r) \O_m (P)$.
The most general form of the connection
is given by
\eqn\ekom{( \d h \cdot \K ) \O_m = \d h (a~ O_m + b~ m)~,}
where $\d h \equiv \d h_\mu^\mu$ is the trace, and $a, b$
are constants. Using the trace condition \etrace, we find
\eqn\eab{a = {1 \over 4}~,\quad b = {1 \over 4}
\left( \cmmone - \bone \right)~,}
where the constant $\cmmone$ is a matrix element
of the connection $c_m$ for the conjugate field $\O_m$.  Under
a redefinition of the cosmological constant
\eqn\eredefgone{\gone \to \gone + {k \over 2}~m^2~,}
the conjugate field $\O_m$ and the connection $\cmmone$
change by
\eqn\eredefOm{\O_m \to \O_m - k m~,\quad
\cmmone \to \cmmone + k~,}
so that the linear combinations
\eqn\ecombinations{\O_m + m \cmmone~, \quad m \O_m + 2 \gone}
are invariant under the redefinition \eredefgone.
In particular, the trace \eIsingtrace\ is invariant.

Taking the column vector $(\O_m, 1)^T$ as the basis,
the general formula \eS\ gives
\eqn\eIsingS{S (\rho;m) = \pmatrix{0&0\cr
m \left[- \bone (\ln \rho + 1) + \cmmone \right]& 1\cr}~.}
Hence, \eCdS\ and \ecomm\ give
\eqnn\eIsingCm
\eqnn\eIsingCtm
$$
\eqalignno{&\d h \cdot \C (\rho; h,m) \O_m ={1 \over 4}~\d h~{ -
m \bone \over \rho}&\eIsingCm\cr
& u \cdot \Ct (\rho; h,m) \O_m \cr
&\quad = u^\mu \p_\mu \O_m
+ {1 \over 2}~\nabla_\mu u^\mu \cdot \left(
\O_m + m \cmmone  - m \bone (1 + \ln \rho)
\right)~. &\eIsingCtm\cr}
$$

Now we consider the product of two energy-momentum tensors.
The most general form of the connection, allowed by covariance
and the $Z_2$ invariance under $m \to - m$, is given as follows:
\eqn\eKgeneral{\eqalign{&\d h \cdot \K (h,m,\gone) \H^\ab\cr
&= {1 \over 2}~ h^\ab \d h \left( A_1 + \gone A_3 + m^2 A_5 +
R A_7 \right)
+ \d h^\ab \left( A_2 + \gone A_4 + m^2 A_6 +
R A_8 \right)\cr
&\quad + {C_1 \over 2}~\d h~h^\ab \H + C_2 \d h \H^\ab +
C_3 h^\ab \d h_\mn \H^\mn + C_4 \d h^\ab \H\cr
&\qquad\qquad + {B_1 \over 4} h^\ab \nabla^2 \d h +
{B_2 \over 4} (\nabla^\alpha \nabla^\beta +
\nabla^\beta \nabla^\alpha) \d h \cr
&\qquad\qquad\qquad\qquad+ {B_3 \over 2}~h^\ab \nabla^\mu
\nabla^\nu \d h_\mn + {B_4 \over 2}~\nabla^2 \d h^\ab~,\cr}}
where $A$'s, $B$'s, and $C$'s are all constants.

The constants $A, B, C$'s will be determined in four steps.  First we use the
integrability constraints obtained in sect.~2 to relate some of the unknown
constants.  Second we will impose consistency with the
previous results \eIsingCm, \eIsingCtm\
on the trace of the energy-momentum tensor.  Third, we will
write down the OPE coefficients $\C, \Ct$ in terms of the
connection $\K$ and impose the algebraic constraints \eCconstraint,
\eCtconstraint.  At this point we will still have some
undetermined constants.
Finally we will determine the remaining constants by
imposing the algebraic constraints
\eCttrace\ and \eCCt.

\subsec{torsion constraint}

The integrability condition \eintegrability\ (or equivalently
\etorsion) gives,
for $\rho = 1$, the following conditions:
\eqnn\etorsioni
\eqnn\etorsionii
$$
\eqalignno{&(\K^\mn) \H^\ab = h^\mn \H^\ab - h^\ab \H^\mn +
(\K^\ab) \H^\mn &\etorsioni\cr
&(\K^{\mn, \gamma\d}) \H^\ab = (\K^{\ab, \gamma\d}) \H^\mn~.
&\etorsionii\cr}
$$
Eq.~\etorsioni\ gives
\eqn\eCtwoCthree{C_3 = C_2 - {1 \over 2}~,}
and eq.~\etorsionii\ gives
\eqn\eBtwoBthree{B_2 = B_3~.}

\subsec{trace condition}

The trace anomaly \eIsingtrace\ provides further
constraints.  Using the explicit form of the trace
anomaly, the variational formula gives
\eqn\evari{\eqalign{&- \vvev{(h+\d h)_\ab \H^\ab (P)}_{h+\d h,g}
+ \vev{h_\ab \H^\ab (P)} \cr
=& ~m \left[ - \vvev{\O_m}_{h+ \d h, g}
+ \vev{\O_m} \right] + c' (- R(h+\d h) + R(h)) \cr
=& ~m \int_{\rho \ge \ep} d^D r \sqrt{h}~{1 \over 2} \d h_\mn (r)
\vev{\H^\mn (r) \O_m (P)}^c \cr
& + m \left[ \d h \cdot \left\{ \K - \int_\ep^1 d\rho~\C (\rho) \right\}
\right]_m^{~a} \vev{\Phi_a (P)} + c' (- R(h+\d h) + R(h))~.\cr}}
On the other hand, the variational formula for the
energy-momentum tensor gives
\eqn\evarii{\eqalign{&- \vvev{(h+\d h)_\ab \H^\ab (P)}_{h+\d h,g}
+ \vev{h_\ab \H^\ab (P)} \cr
= & - \d h_\ab \vev{\H^\ab}
+ m \int_{\rho \ge \ep} d^D r \sqrt{h}~{1 \over 2} \d h_\mn (r)
\vev{\H^\mn (r) \O_m (P)}^c \cr
&+ h_\ab \vev{\left( \d h \cdot \K \right) \H^\ab}
- \int_\ep^1 d\rho~m \left( \d h \cdot \C (\rho)
\right)_m^{~a} \vev{\Phi_a}~.\cr}}
We demand consistency between the two expressions.
Using the variation of the Ricci curvature
\eqn\eRvar{- R(h+\d h) + R(h) = \d h_\mn R^\mn + \nabla^2 \d h
- \nabla^\mu \nabla^\nu \d h_\mn~,}
we obtain
\eqna\etracecondition
$$
\eqalignno{h_\ab (\K^\mn) \H^\ab &= m (\K^\mn) \O_m + 2 c' R^\mn
+ 2 \H^\mn &\etracecondition a\cr
h_\ab (\K^{\mn, \gamma\d}) \H^\ab &= c' \left( 4 h^{\gamma\d} h^\mn
- 2 (h^{\mu\gamma} h^{\nu\d} + h^{\mu\d} h^{\nu\gamma} )
\right)~. &\etracecondition b\cr}
$$
We can obtain \etracecondition{a} also from eq.~\ecKrelation\
of the previous section.
With the connection \ekom, \eab\ for the conjugate field $\O_m$,
eq.~\etracecondition{a} gives
\eqnn\eAs
\eqnn\eCs
$$
\eqalignno{&A_1 + A_2 = 0~,\quad
A_3 + A_4 = - {1 \over 2}~,\cr
&A_5 + A_6 = {1 \over 4} \cmmone - {3 \over 8} \bone~,\quad
A_7 + A_8 = {c' \over 4}~,&\eAs \cr
&C_1 + C_4 = - {3 \over 4}~,\quad
C_2 = 1~,\quad C_3 = {1 \over 2}~,&\eCs\cr}
$$
and eq.~\etracecondition{b} gives
\eqn\eBs{B_1 + B_4 = 3 c'~,\quad
B_2 = B_3 = - c'~.}
We have used eqs.~\eCtwoCthree\ and \eBtwoBthree.

\subsec{OPE coefficients $\C, \Ct$ in terms of the connection $\K$}

Using eqs.~\eSH\ and \eSKH, we can write down the OPE
coefficients $\C$ in terms of the connection
\eKgeneral\ as follows:
\eqna\eCintermed
$$
\eqalignno{(\C^\mn (\rho)) \H^\ab &= - {2 \over \rho^3}
\left[ h^\mn h^\ab A_1 + (h^{\mu\alpha} h^{\nu\beta} +
h^{\mu\beta} h^{\nu\alpha} ) A_2 \right] \cr
+ &{\bone m^2 \over 2 \rho}
\left[ h^\mn h^\ab A_3 + (h^{\mu\alpha} h^{\nu\beta} +
h^{\mu\beta} h^{\nu\alpha} ) A_4 \right] &\eCintermed a\cr
(\C^{\mu, \gamma\d} (\rho)) \H^\ab &= 0~.&\eCintermed b\cr}
$$
Using the algebraic constraint \eCconstraint, we find that
eq.~\eCintermed{b} implies the vanishing of $A_1$, $A_2$:
\eqn\eAoneAtwo{A_1 = A_2 = 0~.}
This is consistent with \eAs.

Using the general formula \ecomm, we can calculate the
coefficients $\Ct$.  The result is as follows:
\eqna\eCtintermed
$$
\eqalignno{&(\Ct^{\mn,}_{~~~~\mu} (\rho)) \H^\ab \cr
&= {1 \over 6} \left[
(- B_1 - c' - 2 B_4) h^\ab \p^\nu R +
(c' + B_4) (h^{\nu\beta} \p^\alpha R +
h^{\nu\alpha} \p^\beta R) \right]\cr
&\quad + \nabla^\nu \H^\ab~, &\eCtintermed a\cr
&(\Ct^{\mn ,~~\gamma}_{~~~~\mu} (\rho)) \H^\ab \cr
&= h^{\gamma\nu} h^\ab \Big[ A_3 (\gone + {1 \over 2}
\bone m^2 \ln \rho ) + m^2 A_5 \cr
&\qquad + R (A_7
+ {1 \over 6} (- B_1 -2 c' - 3 B_4)) + C_1 \H \Big]\cr
&\quad + (h^{\nu\alpha} h^{\gamma\beta} + h^{\nu\beta} h^{\gamma\alpha})
\Big[ A_4 (\gone + {1 \over 2} \bone m^2 \ln \rho) + m^2 A_6 \cr
&\qquad + R ( A_8 + {1 \over 3} ( c' + B_4 )) + C_4 \H \Big] \cr
&\quad - h^{\nu\alpha} \H^{\beta \gamma} - h^{\nu\beta}
\H^{\alpha \gamma} + 2 h^{\gamma\nu} \H^\ab + h^\ab
\H^{\nu\gamma}~, &\eCtintermed b\cr
&(\Ct^{\mn ,~~\gamma\d\ep}_{~~~~\mu} (\rho)) \H^\ab
= (\K^{\nu\gamma, \d\ep} + \K^{\nu\d, \ep\gamma} + \K^{\nu\ep,
\gamma\d}) \H^\ab \cr
&= (B_1 - c') g^\ab (h^{\nu\gamma} h^{\d\ep} + h^{\nu\d}
h^{\ep\gamma} + h^{\nu\ep} h^{\gamma\d}) \cr
&\quad - c' \left(
h^{\nu\gamma} ( h^{\d\alpha} h^{\ep\beta}
+ h^{\d\beta} h^{\ep\alpha}) + ...~ \right)\cr
&\quad \quad+ B_4 \left( h^{\gamma\d} (h^{\ep\alpha} h^{\nu\beta}
+ h^{\ep\beta} h^{\nu\alpha}) + ...~ \right)~,
&\eCtintermed c\cr}
$$
where we have used eqs.~\eCs\ and \eBs,
and the last omitted terms are obtained by symmetrizing with
respect to $\gamma$, $\d$, and $\ep$.

Since eq.~\eCtintermed{b} has no $1/\rho^2$ singularity, the
algebraic constraint \eCtconstraint\ implies that
\eqn\eCtvanishing{h_{\d\ep} (\Ct^{\mn,~~\gamma\d\ep}_{~~~~\mu})
\H^\ab = 0~.}
Using \eBs, this gives
\eqn\eBoneBfour{B_1 = {5 c' \over 2}~,\quad B_4 = {c' \over 2}~.}

We also notice that the right-hand side of eq.~\eCtintermed{b}
should not depend on the cosmological constant $\gone$,
since the short-distance singularities are independent of
the additive constant in the vacuum energy.  Using the trace
anomaly \eIsingtrace, the absence of $\gone$ in
eq.~\eCtintermed{b} gives
\eqn\eAthreeAfour{A_3 = - 3 - 2 C_1~,\quad
A_4 = {5 \over 2} + 2 C_1~.}

To summarize so far, eqs.~\eAoneAtwo, \eAthreeAfour,
\eBs, \eBoneBfour, and \eCs\ give all the constants
except for $A_6$, $A_8$, and $C_1$.  To determine
these three remaining constants, we must use the algebraic constraints
\eCttrace\ and \eCCt.

\subsec{Further algebraic constraints}

To determine the remaining unknown constants, we
try to construct coefficients $\Ct^{\mn, \mu_1 ... \mu_m}$ which
are related to eqs.~\eCtintermed{} by
eqs.~\eCttrace.  At the same time we must satisfy the constraints
\eCCt.  We have found it convenient to do this construction in a complex
coordinate system $z$ in which the metric has only $h_{z\zbar}$
non-vanishing.  While constructing the coefficients
$\Ct^{\mn, \mu_1 ... \mu_m}$, it is also important to satisfy
the Bose symmetry \einterchangeH.  We will omit the detail
here.  The final results are as follows:
\eqn\efinal{\eqalign{&A_3 = A_4 = - {1 \over 4}~,\quad
A_5 = A_6 = {1 \over 8} \left( \cmmone  - {3 \over
2} \bone \right)~,\cr
&A_7 = {7 c' \over 8}~,\quad A_8 = - {5 c' \over 8}~,\quad
C_1 = - {11 \over 8}~,\quad C_4 = {5 \over 8}~.\cr}}

Written in a more transparent form, our final results are
given by
\eqna\eresult
$$
\eqalignno{&{1 \over \rho} \int_{\rho (z,P) = \rho} d
\Omega_\rho (z,\zbar)~\d h_\mn \H^\mn (z,\zbar) \H^{zz} (P)\cr
&=- {c' \over 4} (\nabla^z)^4 \d h_{zz} (P) +
\nabla^z \d h_{zz} \cdot \nabla^z \H^{zz} (P) + \nabla^z
\nabla^z \d h_{zz} \cdot \H^{zz} (P) \cr
&\quad - {m^2 \bone \over 16} ~ \nabla^2 \d h^{zz} (P) &\eresult a \cr
&\quad + {m \over 4} \left( - \nabla^z \d h \cdot \nabla^z - \nabla_z \d
h^{zz} \cdot \nabla^z + \nabla^z \d h^{zz} \cdot \nabla_z
\right) \O_m (P) + {\rm o} (\rho^0)~,\cr
&{1 \over \rho} \int_{\rho (z,P) = \rho} d \Omega_\rho (z,\zbar)~\d h_\mn
\H^\mn (z,\zbar) \O_m (P)\cr
&= {1 \over 2} \left( \nabla^z \nabla^z \d h_{zz} + \nabla_z
\nabla_z \d h^{zz} \right) \left[ {1 \over 2}
(\O_m (P) + m \cmmone ) - m \bone \left( {1 \over 4} + {1 \over 2}
\ln \rho \right) \right] \cr
&\quad - {m \bone \over 8} ~\nabla^2 \d h (P) &\eresult b\cr
&\quad + \left( \nabla^z \d h_{zz} \cdot \nabla^z +
\nabla_z \d h^{zz} \cdot \nabla_z \right) \O_m (P) + {\rm o} (\rho^0)~,\cr}
$$
where $\d h_\mn$ is an arbitrary symmetric tensor which vanishes
at $P$.  In the $m=0$ limit, this agrees with conformal
field theory \ref\rbpz{A.~A.~Belavin, A.~M.~Polyakov, and
A.~B.~Zamolodchikov, Nucl.~Phys.~{\bf B241}(1984)333}.

\newsec{Conclusion}

In this paper we have continued the study of the energy-momentum
tensor initiated in ref.~\rI.  In particular we have examined
integrability of the first order variational formula
\efirstorder\ for the vacuum energy, and derived the condition
\eintegrability.   We have observed that this integrability
condition \eintegrability\ gives the torsion of the connection
$\K$ up to total derivatives as in eq.~\etorsion.
We have also interpreted the
dependence of eq.~\eintegrability\
on the geodesic distance $\rho$ as the Bose symmetry between two
energy-momentum tensors.  Finally, using the integrability
condition and other algebraic constraints,
we have determined the short-distance singularities of the
product of two energy-momentum tensors in the massive Ising
model in two dimensions.

The massive Ising model may be too simple.  After all, it is
a theory of the free massive Majorana fermion, and the OPE of
two energy-momentum tensors can be calculated directly using
the elementary spinor fields (at least in flat space).
The aim of the example is to show that we understand the general
properties of the energy-momentum tensor well enough to
determine the OPE explicitly without explicit calculations.

The goal of ref.~\rI\ and the present paper is understanding
the first order variational formula \evar, which actually
defines the energy-momentum tensor through its integral over
space.  We have applied only a limited integrability check
in this paper.  In a separate paper \ref\rIII{H.~Sonoda,
``The Energy-Momentum Tensor in Field Theory III,'' paper in
preparation} we plan to address the full integrability condition
for the variational formula applied to arbitrary correlation
functions instead of the vacuum energy.

\listrefs

\bye